\documentclass[10pt,conference]{IEEEtran}
\IEEEoverridecommandlockouts

\usepackage[style=ieee,minnames=1,maxcitenames=2]{biblatex} 
\usepackage{amsmath,amssymb,amsfonts}
\usepackage[inline]{enumitem}
\usepackage{algorithm}
\usepackage{algpseudocode}
\usepackage{graphicx}
\usepackage{textcomp}
\usepackage[dvipsnames]{xcolor}
\usepackage{subcaption}
\usepackage{multirow}
\usepackage{siunitx}
\usepackage{comment}

\mathchardef\mhyphen="2D 

\algnewcommand{\LineComment}[1]{\Statex \(\triangleright\) #1}

\begin{document}

\renewcommand{\arraystretch}{1.25}

\title{Explaining Software Bugs Leveraging Code Structures in Neural Machine Translation
}

\author{
\IEEEauthorblockN{Parvez Mahbub}
\IEEEauthorblockA{\textit{Department of Computer Science} \\
\textit{Dalhousie University}\\
Nova Scotia, Canada \\
parvezmrobin@dal.ca}
\and
\IEEEauthorblockN{Ohiduzzaman Shuvo}
\IEEEauthorblockA{\textit{Department of Computer Science} \\
\textit{Dalhousie University}\\
Nova Scotia, Canada \\
oh599627@dal.ca}
\and
\IEEEauthorblockN{Mohammad Masudur Rahman}
\IEEEauthorblockA{\textit{Department of Computer Science} \\
\textit{Dalhousie University}\\
Nova Scotia, Canada \\
masud.rahman@dal.ca}
}

\maketitle

\begin{abstract} 
Software bugs claim $\approx$~50\% of development time and cost the global economy billions of dollars.
Once a bug is reported, the assigned developer attempts to identify and understand the source code responsible for the bug and then corrects the code. 
Over the last five decades, there has been significant research on automatically finding or correcting software bugs. However, there has been little research on automatically explaining the bugs to the developers, which is essential but a highly challenging task.
In this paper, we propose Bugsplainer, a transformer-based generative model, that generates natural language explanations for software bugs by learning from a large corpus of bug-fix commits. Bugsplainer can leverage structural information and buggy patterns from the source code to generate an explanation for a bug.
Our evaluation using three performance metrics shows that Bugsplainer can generate \emph{understandable} and \emph{good} explanations according to Google's standard, and can outperform multiple baselines from the literature.
We also conduct a developer study involving 20 participants where the explanations from Bugsplainer were found to be more accurate, more precise, more concise and more useful than the baselines.
\end{abstract}


\section{Introduction}
\label{sec: introduction}
A software bug is an incorrect step, process, or data definition in a computer program that prevents the program from producing the correct result~\parencite{ieee-standard}.
Bug resolution is one of the major tasks of software development and maintenance. According to several studies, it consumes up to 40\% of the total budget~\parencite{Glass2001} and costs the global economy billions of dollars each year~\parencite{Britton2013, zou2018practitioners}.

When an end-user reports a software bug, the assigned developer attempts to identify and understand the source code responsible for the bug and then corrects the code.
Over the last five decades, there have been numerous approaches to automatically find the location of a bug~\parencite{zou2018practitioners, rahman2021forgotten}.
However, they often identify certain parts of the code as buggy without offering any meaningful explanation~\parencite{kochhar2016practitioners}.
Developers are thus generally responsible for understanding a bug from the identified code before making any changes. Understanding bugs by looking at the code claims a significant chunk of debugging time. In fact, developers spend $\approx$~50\% of their time comprehending the code during software maintenance~\parencite{roehm2012professional}.
However, neither many studies attempt to explain the bugs in the source code to the developers, nor are they practical and scalable enough for industry-wide use~\parencite{kochhar2016practitioners, zou2018practitioners}.

Explaining a bug in the software code is essential to fix the bug, but a highly challenging task.
Many static analysis tools such as FindBugs~\parencite{FindBugs}, PMD, SonarLint, PyLint, and pyflakes~\parencite{pyflakes} employ complex hand-crafted rules to detect the bugs and vulnerabilities in software code.
Upon detection, they use pre-defined message templates to explain the bugs and vulnerabilities.
Unfortunately, their utility could be limited due to their high false-positive results and the lack of actionable insights in their explanations~\parencite{barik2016should, johnson2013don, ayewah2007evaluating}.
In particular, their explanations are often too generic and unaware of the context due to their pre-defined, templated nature~\parencite{nachtigall2022large}.
\textcite{thung2015extent} also suggest that static analysis tools suffer from a large number of \emph{false negative} results, which could leave the software systems vulnerable to bugs.

Unlike traditional, rule-based approaches (e.g., static analysis tools), explaining software bugs can be viewed as a translation task, where the buggy code is the source language and the corresponding explanation is the target language.
In recent years, machine translation, especially neural machine translation (NMT)~\parencite{jurafsky_martin_2014}, has found numerous applications in several domains~\parencite{devlin2018bert, raffel2020exploring}.
NMT has also been used in different software engineering tasks including, but not limited to, code summarization~\parencite{loyola2017neural, hu2018deep}, code comment generation~\parencite{tufano2021towards, hong2022commentfinder}, and commit message generation~\parencite{jiang2017automatically, tao2021evaluation, liu2020atom, liu2018neural}.
Traditional NMT models often consist of two items: \emph{encoder} and \emph{decoder}.
The encoder first converts the words of the source language into an intermediate numeric representation.
Then the decoder generates the target words one by one using the intermediate representation and previous words from the generated sequence~\parencite{jurafsky_martin_2014}.
However, explanation generation from the buggy source code using neural machine translation poses two major challenges as follows.

\begin{figure*}
\centering
\begin{minipage}{0.55\textwidth}
    \centering
    \includegraphics[width=\linewidth]{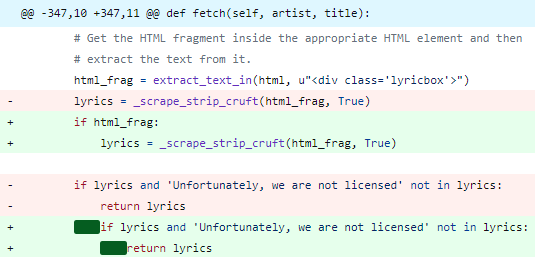}
    \caption{An example of buggy source code}%
    \label{fig: motivating-example}%
\end{minipage}
\begin{minipage}{0.44\textwidth}
    \centering
    \captionsetup{type=table} 
    \caption{Generated explanations for buggy code}%
    \label{tab: generated-explnations}%
    \footnotesize
    \begin{tabular}{|p{18mm}|p{45mm}|}
        \hline
        \textbf{Technique} & \textbf{Generated Explanation} \\
        \hline
        \hline
        Ground Truth &
        Fix a bug where the lyricswiki fetcher would  try to unescape an empty (None) response and crash  \\
        \hline
        CommitGen~\parencite{jiang2017automatically} & Small bug fix for error handling \\
        \hline
        NNGen~\parencite{liu2018neural} & fix UnicodeDecodeError with non-ASCII text \\
        \hline
        Fine-tuned \newline CodeT5~\parencite{wang2021codet5} & Don't try to get lyrics if we are licensed \\
        \hline
        pyflakes~\parencite{pyflakes} & \emph{no error found} \\
        \hline
        \hline
        \textbf{Bugsplainer} & fix crash when lyrics not found \\
        \hline
    \end{tabular}
\end{minipage}
\end{figure*}

\textbf{Understanding the structures of source code}:
Natural language is loosely structured, which exhibits phenomena like ambiguity and word movement~\parencite{jurafsky_martin_2014}.
Word movement is the appearance of words in a sentence in different orders but still being grammatically correct.
On the contrary, programming languages are more structured, syntactically restricted, and less ambiguous~\parencite{allamanis2018survey}.
From the two programs having the same vocabulary, one could be buggy and the other could be correct due to their structural differences (e.g., \ref{subfig: example-buggy}, \ref{subfig: example-bug-free}).
Thus, capturing and understanding the structures of code is essential to explaining the buggy code.
Unfortunately, traditional NMT-based techniques often treat source code as a sequence of tokens and thus might fail to capture the structures of source code properly~\parencite{iyer2016summarizing}.

\textbf{Understanding and detecting buggy code patterns}: From a high-level perspective, NMT models translate words from the source language into words from the target language. 
However, to generate explanations from the buggy code, the model must be able to accurately reason about the bug from the buggy code and its structures. 
Such reasoning is non-trivial and warrants the model to be aware of buggy code patterns. 
Traditional NMT models might not be sufficient to tackle all these challenges due to their simplified assumptions about sequential inputs and outputs. 
According to~\textcite{ray2016naturalness} buggy code is less repetitive (a.k.a., \textit{unnatural}) than regular code, which could exacerbate the above challenges.

In this paper, we propose \emph{Bugsplainer}, a novel transformer-based generative model, that generates natural language explanations for software bugs by learning from a large corpus of bug-fix commits (i.e., commits that correct bugs).
Our solution is able to address the above challenges, which makes our work \emph{novel}. 
First, Bugsplainer can leverage code structures in explanation generation by applying structure-based traversal~\parencite{hu2018deep} to the buggy code.
Second, we train Bugsplainer using both buggy source code and its corrected version, which helps the model to understand and detect buggy code patterns during its explanations generation for the buggy code.

We train and evaluate Bugsplainer with $\approx$~150K bug-fix commits collected from GitHub using three different metrics -- BLEU~\parencite{papineni2002bleu}, Semantic Similarity~\parencite{haque2022semantic} and Exact Match.
We find that the explanations from Bugsplainer are \emph{understandable} to \emph{good}.
We compare our technique with four appropriate baselines -- pyflakes~\parencite{pyflakes}, CommitGen~\parencite{jiang2017automatically}, NNGen~\parencite{liu2018neural}, and Fine-tuned CodeT5~\parencite{wang2021codet5}.
Bugsplainer outperforms all four baselines in all metrics by a statistically significant margin.
One major strength of Bugsplainer is understanding the structure of the code and buggy code patterns, where the baselines might be falling short.
To further evaluate our work, we conduct a developer study involving 20 developers from six countries, where the identities of both our tool and the baselines were kept hidden.
The study result shows that explanations from Bugsplainer are more accurate, more precise, more concise, and more useful compared to that of the baselines.

We thus make the following contributions in this paper: 

\begin{enumerate}[label=(\alph*)]
    \item A novel transformer-based technique, \emph{Bugsplainer}, 
    that can explain software bugs by leveraging the structural information and buggy code patterns from source code.
    \item A novel pre-training technique, namely -- Discriminatory Pre-training, that is shown to be effective in generating better explanations.
    \item A benchmark dataset containing $\approx$~150K instances of buggy code, corrected code, and corresponding explanations written by human developers. To the best of our knowledge, this is the first benchmark of its kind.
    \item A comprehensive evaluation and validation of the Bugsplainer technique using both popular performance metrics (e.g., BLEU score) and a developer study. 
    \item A replication package that includes our working prototype, experimental dataset, and other configuration details for the replication or third-party reuse\footnote{https://bit.ly/3H7R1aI}.
\end{enumerate}

\section{Motivating Example}
\label{sec: motivating-example}

To demonstrate the capability of our technique -- \emph{Bugsplainer}, let us consider the example in Fig.~\ref{fig: motivating-example}.
The code snippet is taken from \texttt{beetbox/beets} repository at GitHub\footnote{https://bit.ly/3PGnkzK}.
The buggy code attempts to scrape the lyrics of a song from an HTML fragment.
However, if the HTML fragment is empty, then the program crashes.
Table~\ref{tab: generated-explnations} shows both developer's explanation for the buggy code (a.k.a., reference) and the explanations generated by different techniques including Bugsplainer.
We see that the explanations generated by CommitGen~\parencite{jiang2017automatically} (i.e., RNN-based technique) and NNGen~\parencite{liu2018neural} (i.e., Information Retrieval-based technique) are not helpful.
On the other hand, the explanation from Fine-tuned CodeT5~\parencite{wang2021codet5} is not accurate as the bug has nothing to do with licensing.
Pyflakes, a static analysis-based technique, does not provide any explanation since it was not able to detect the bug using its pre-defined rules.
On the other hand, the explanation generated by our technique, Bugsplainer, is \emph{accurate} as it expresses the same information as the ground truth and \emph{precise} as it expresses no unnecessary information.
Moreover, we see that in the fixed version of the code (Fig.~\ref{fig: motivating-example}), an \textbf{\texttt{if}} condition was used to check whether the HTML fragment exists (i.e., lyrics were found) or not, which reflects the solution implied by our explanation.

\section{Background}
\label{sec: background}
\subsection{Neural Machine Translation}
Neural machine translation (NMT) is a deep neural network-based approach for automated translation~\parencite{wu2016google}.
In recent years, NMT has achieved rapid progress and has drawn the attention of both the research community and the practitioners.
Generally, an NMT model is composed of two different blocks: \emph{encoder} and \emph{decoder}.
The encoder accepts an input sequence and produces a numerical, intermediate representation of the input.
Then, this intermediate representation is passed to the decoder. Based on this intermediate representation, the decoder starts to generate the target sequence, one token at a time.
While generating each token, all the previously generated tokens are also passed to the decoder.
This is known as \emph{autoregressive process} where the current output is based on all previously generated outputs~\parencite{vaswani2017attention}.
\textcite{bahdanau2014neural} demonstrate how certain locations of the input sequence can be emphasized over others for an effective translation, which is known as the attention mechanism.
The attention mechanism makes the training process faster and helps the NMT models translate long sequences~\parencite{vaswani2017attention}.
In our research, we use Transformer~\parencite{vaswani2017attention, raffel2020exploring}, the state-of-the-art NMT model along with the attention mechanism, as a part of Bugsplainer, to generate explanations for the buggy source code.

\subsection{Structure-Based Traversal}
\label{subsec: sbt}

\begin{figure}
    \centering
    \includegraphics[width=0.7\linewidth]{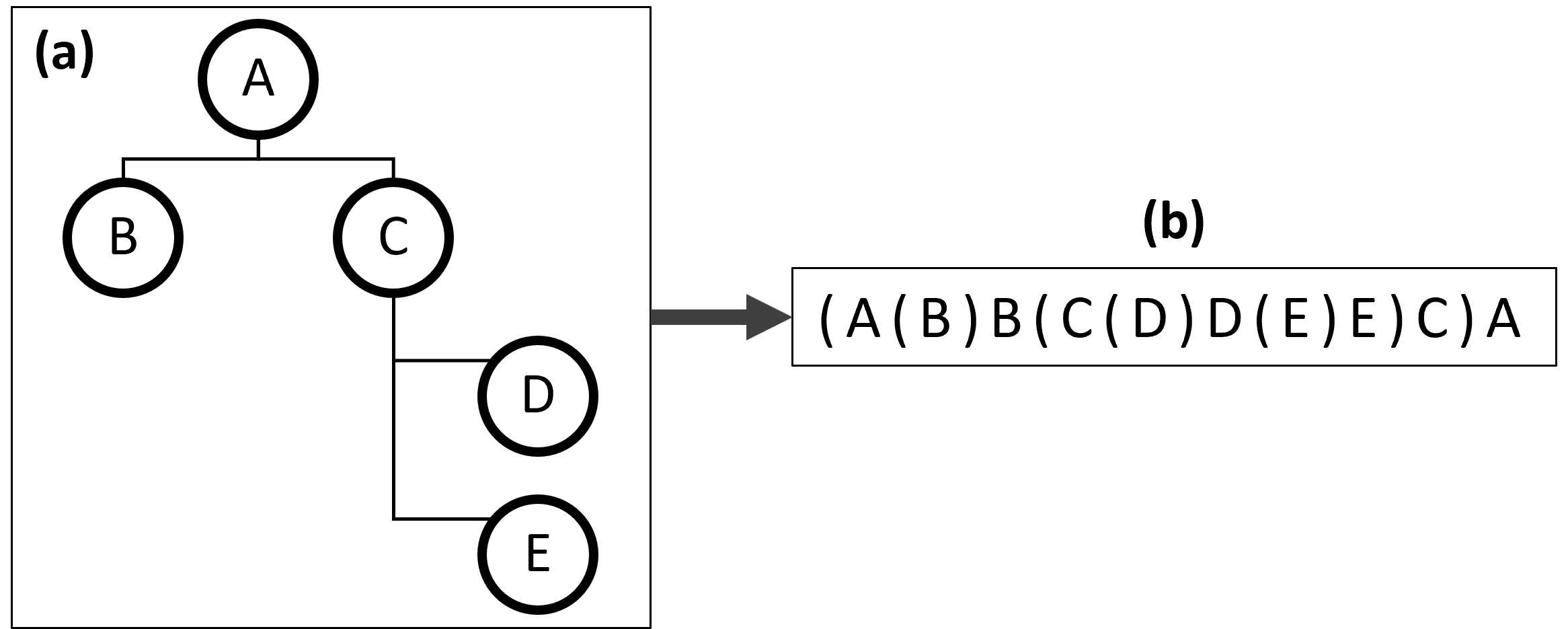}
    \caption{Structure-based traversal (SBT) -- (a) an example tree, and (b) corresponding SBT sequence}
    \label{fig: sbt-numbers}
\end{figure}

Traditional NMT models treat their input as sequential data (e.g., English language texts).
However, source code is rich in structures (e.g., syntactic or data dependencies), which are essential to convey the semantics of the code.
One way to represent the syntactic structure of a source code document is an abstract syntax tree (AST).
To leverage this structural information, several studies represent the tree structure into a sequence of code tokens and use it as the input to sequence-to-sequence models~\parencite{hu2018deep, liu2020atom}.
\textcite{hu2018deep} propose structure-based traversal (hereby SBT) to convert an AST node into a token sequence that can preserve the structural information.
Fig.~\ref{fig: sbt-numbers} shows an example tree and its corresponding SBT sequence. We use the SBT algorithm of \citeauthor{hu2018deep} to generate the sequence (see Algorithm~\ref{algo: diff-sbt} for details).

To generate the SBT sequence of a tree, we first use a pair of brackets to represent the tree structure and put the root node (i.e., \texttt{A}) behind the right bracket, i.e., \texttt{(A)A}.
Next, we traverse the sub-trees of the root node and put all root nodes of sub-trees into the brackets, i.e., \texttt{(A(B)B(C)C)A}.
Recursively, we traverse each sub-tree until all nodes of a tree are traversed. 
For example, we get the following SBT sequence --\texttt{(A(B)B(C(D)D(E)E)C)A} -- for the example tree in Fig.~\ref{fig: sbt-numbers}a.

\section{Bugsplainer}
\label{sec: methodology}

\begin{figure*}
    \centering
    \begin{subfigure}{\linewidth}
        \centering
        \includegraphics[width=.9\linewidth]{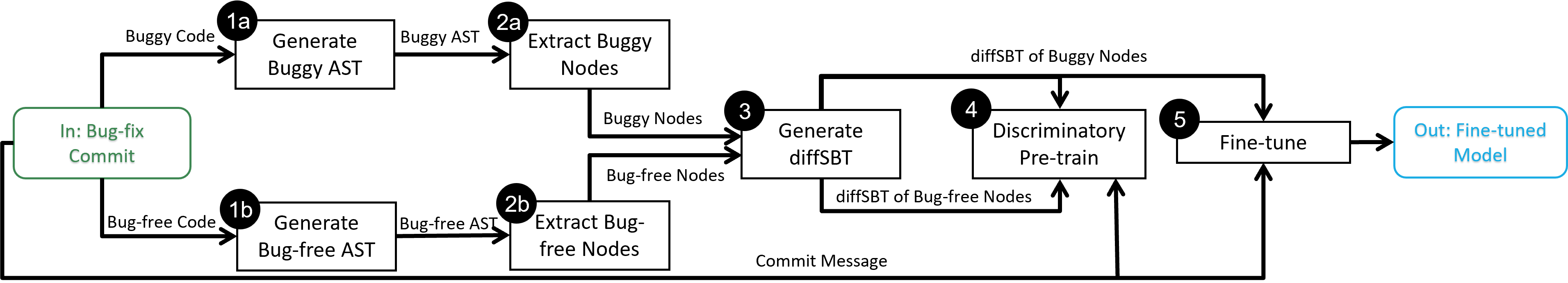}
        \caption{Training of Bugsplainer}
        \label{subfig: schema-training}
    \end{subfigure}
    \hfill%
    \vspace{0.5em}
    \begin{subfigure}{\linewidth}
        \centering
        \includegraphics[width=.85\linewidth]{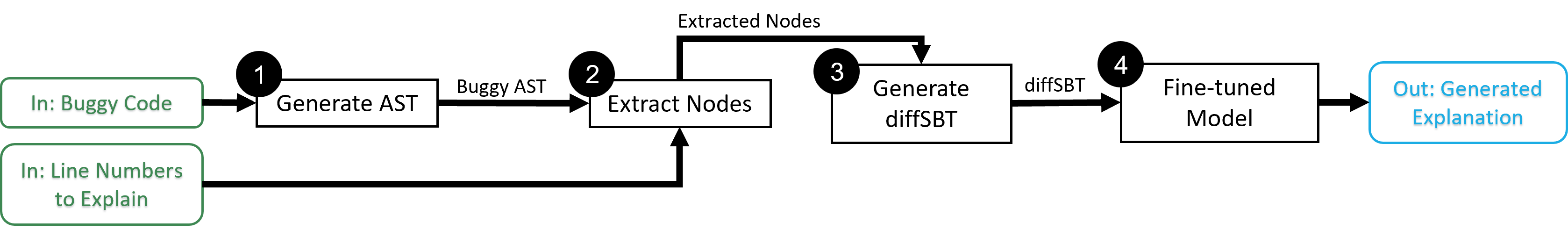}
        \caption{Explanation generation for buggy code}
        \label{subfig: schema-testing}
    \end{subfigure}
    \caption{Schematic diagram of Bugsplainer}
    \label{fig: schematic-diagram}
\end{figure*}

Fig.~\ref{fig: schematic-diagram} shows the schematic diagram of our proposed technique -- \emph{Bugsplainer} -- for explaining software bugs.
We discuss different steps of our technique in detail as follows.

\subsection{Extract Buggy and Bug-free AST Nodes from Commit}
\label{subsec: extract-nodes}
First, we construct abstract syntax trees (AST) of both buggy and bug-free code using the information from a bug-fix commit.
A bug-fix commit contains the bug-free version of the code while being connected to its parent commit containing the buggy version.
From these two versions of the source code, we construct two different ASTs -- the buggy AST (Step 1a, Fig.~\ref{subfig: schema-training}) and the bug-free AST (Step 1b, Fig.~\ref{subfig: schema-training}).
A commit also contains references to both removed lines (i.e., buggy lines) and added lines (i.e., bug-fix lines).
Using these line numbers, we extract the buggy nodes from the buggy AST (Step 2a, Fig.~\ref{subfig: schema-training}) and bug-free nodes from the bug-free AST (Step 2b, Fig.~\ref{subfig: schema-training}).
If a multi-line expression touches these line numbers, we extract the whole expression node (Line~\ref{diffsbt: is-expression}, Algorithm~\ref{algo: diff-sbt}).
Besides the affected lines, the contextual information (e.g., surrounding lines) often provides useful clues about why the code was changed.
\textcite{asaduzzaman2014cscc} suggest that three lines of code around a target line might be sufficient to capture the contextual information.
While extracting the buggy and bug-free nodes, we thus also extract the nodes representing three lines above and below the changed lines in the code.

\setlength{\textfloatsep}{2pt}
\begin{algorithm}[t]
    \small
    \caption{Generate diffSBT sequence from commit diff}
    \label{algo: diff-sbt}
    
    \begin{algorithmic}[1]
    \Function{diffSBT}{c}
        \Comment{Generate diffSBT sequence for commit}
        \State buggyAST $\gets$ \Call{BuildAST}{c.buggyCode}
        \State bugfreeAST $\gets$ \Call{BuildAst}{c.bugFreeCode}
        \State buggyNodes $\gets$ \Call{Intersections}{buggyAST, c.removed}
        \State bugfreeNodes $\gets$ \Call{Intersections}{bugfreeAST, c.added}
        \State \Return \Call{SBT}{buggyNodes} + $\langle /s\rangle$ + \Call{SBT}{bugfreeNodes}
    \EndFunction
    \Statex
    
    \Function{Intersections}{r, ln}
    
    \State nodes $\gets \phi$
    \Comment{Initialize nodes with an empty list}
    \ForAll{n in r}
    \Comment{Get intersections for all nodes in $r$}
    
        \If{\Call{IsInside}{n, ln} \textbf{or} \Call{IsExpression}{n}}
            \label{diffsbt: is-expression}
            \State \Call{Append}{nodes, n}
            
        \ElsIf{\Call{StartsInside}{r, ln}}
            \Comment{Keep node $r$ but prune the children outside $ln$}
            \State r.children $\gets$ \Call{Intersections}{r.children}
            \State \Call{Append}{nodes, n}
        
        \ElsIf{\Call{EndsInside}{r, ln}}
        \Comment{Node $r$ starts before the $ln$. Return only the children of $r$ that intersect with $ln$.}
            \State children $\gets$ \Call{Intersections}{r.children}
            \State \Call{Append}{nodes, children}
        \EndIf
    
    \EndFor
    
    \State \Return nodes
    
    \EndFunction

    \end{algorithmic}
\end{algorithm}

\subsection{Generate diffSBT Sequence}
\label{subsec: gen-diffsbt}

In this step, we convert the buggy and bug-free AST nodes into diffSBT sequences (i.e., preserve structural information) using the diffSBT algorithm (Step 3, Fig.~\ref{subfig: schema-training}). 
Algorithm~\ref{algo: diff-sbt} shows our algorithm -- \emph{diffSBT} -- for structure-preserving sequence generation from commit diff, which is an adaptation of SBT algorithm by \textcite{hu2018deep}.
We create two versions of the diffSBT sequence.
One of them contains both buggy and bug-free nodes to aid the discriminatory pre-training (see Section~\ref{sssec: discriminatory-pre-training}).
The other contains only buggy nodes to aid the fine-tuning.

\begin{figure*}
    \centering
    \begin{subfigure}[b]{.25\linewidth}
        \centering
        \includegraphics[width=\linewidth]{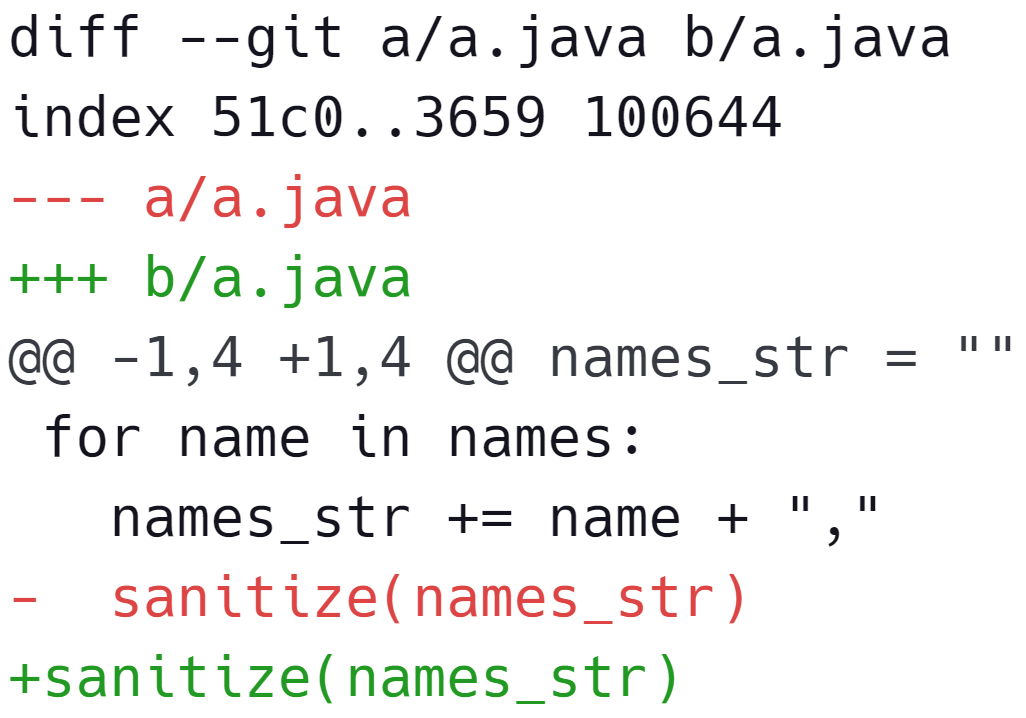}
        \caption{Bug-fix commit diff}
        \label{subfig: example-diff}
    \end{subfigure}
    \hfill%
    \begin{subfigure}[b]{.20\linewidth}
        \begin{subfigure}[b]{\linewidth}
            \includegraphics[width=\linewidth]{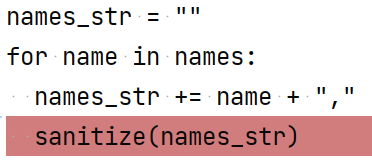}
            \caption{Buggy code}
            \label{subfig: example-buggy}
            \vspace{0.5em}
        \end{subfigure}
        \begin{subfigure}[b]{\linewidth}
            \includegraphics[width=\linewidth]{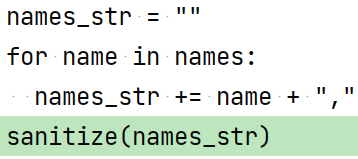}
            \caption{Bug-free code}
            \label{subfig: example-bug-free}
        \end{subfigure}
    \end{subfigure}
    \hfill%
    \begin{subfigure}[b]{.535\linewidth}
        \centering
        \includegraphics[width=\linewidth]{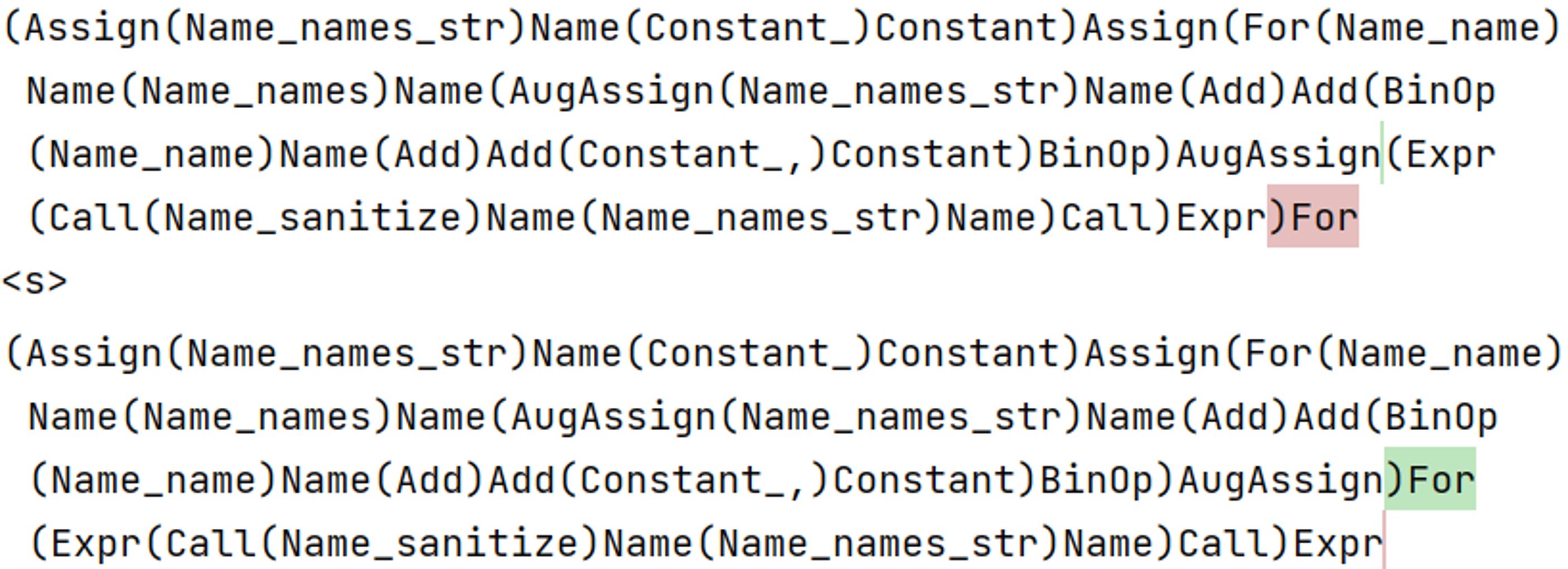}
        \caption{diffSBT sequence for the buggy and bug-free code}
        \label{subfig: example-diffsbt}
    \end{subfigure}

    \caption{An example of diffSBT sequence generation from buggy code and commit diff}
    \label{fig: diff-sbt}
\end{figure*}

To illustrate the generation of diffSBT sequences from a commit diff, let us consider Fig.~\ref{fig: diff-sbt}.
Fig.~\ref{subfig: example-diff} contains a bug-fix commit.
The source code before submitting this commit was buggy (Fig.~\ref{subfig: example-buggy}), and the source code after the commit is bug-free (Fig.~\ref{subfig: example-bug-free}).
The bug is that the \texttt{sanitize()} function was called inside the \texttt{for} loop rather than outside the \texttt{for} loop. 
The buggy code resembles the fixed version of the code as both code segments also have the same vocabulary.
However, they differ by white spaces, as shown in the commit diff (Fig.~\ref{subfig: example-diff}), which is a scope-related problem according to Python programming language.
Thus, if the source code is simply considered as a sequence of tokens without the structural information  (as many studies~\parencite{jiang2017automatically, liu2018neural} do), the bug is really hard to understand.
Fig.~\ref{subfig: example-diffsbt} shows the diffSBT sequence for this bug-fix change.
We see that, in the buggy version, the \texttt{For} block closes at the end of the diffSBT sequence. 
On the contrary, the \texttt{For} block closes before the \texttt{Expr} block in the bug-free version.
Such a placement ensures that the \texttt{Expr} block (i.e., \texttt{sanitize(name\_str)}) is outside the \texttt{For} block.
Thus, with the help of diffSBT, Bugsplainer can identify the difference in the structure of code, which could be useful to explain the bug.

\subsection{Train Bugsplainer}
\label{subsec: training}

In Fig.~\ref{subfig: schema-training}, Steps 4-5 explain the training of Bugsplainer.
Our training phase is divided into two steps -- discriminatory pre-training and fine-tuning.
In both steps, we use a RoBERTa tokenizer~\parencite{liu2019roberta}, pre-trained on GitHub CodeSearchNet dataset~\parencite{husain2019codesearchnet}.
Due to its pre-train dataset (CodeSearchNet), this RoBERTa tokenizer has common code elements in its vocabulary, which can reduce the length of tokenized code sequence by 30\%-45\%~\parencite{wang2021codet5}.
We use this tokenizer to tokenize and encode the inputs (e.g., diffSBT sequence) and decode the outputs (e.g., commit message).
In the following sections, we describe the training phase in detail.

\subsubsection{Discriminatory Pre-training}
\label{sssec: discriminatory-pre-training}

Pre-trained language models have been found to be effective in improving many natural language understanding tasks (e.g., news title generation, question-answering)~\parencite{devlin2018bert, raffel2020exploring}.
During pre-training, a model acquires a general knowledge about a domain which allows it to \emph{understand} the input (e.g., text, image)~\parencite{raffel2020exploring}.
In natural language processing, pre-training is often performed in an unsupervised fashion (e.g., Word2Vec~\parencite{mikolov2013efficient}, missing token prediction~\parencite{devlin2018bert}).
However, many domains use supervised pre-training as well (e.g., Multi-Task Learning~\parencite{raffel2020exploring, wang2021codet5}).
Bugsplainer uses both unsupervised and supervised pre-training to equip the model with a comprehensive understanding of the programming language and its bugs.

We use a pre-trained model -- CodeT5~\parencite{wang2021codet5} -- to perform our discriminatory pre-training with buggy and bug-free code.
CodeT5 is a transformer model based on the Text to Text Transfer Transformer (T5) architecture~\parencite{vaswani2017attention, raffel2020exploring}. 
It has two versions -- 60M parameters and 220M parameters.
We use the 60M parameter version for Bugsplainer, which is pre-trained on GitHub CodeSearchNet data~\parencite{husain2019codesearchnet} for three unsupervised tasks.
CodeSearchNet contains $\approx$~6M methods written in popular programming languages accompanied by natural language documentation.
Thus, the CodeT5 model has a significant understanding of both programming and natural languages, which makes it an ideal choice for our pre-training task.

The pre-training with the CodeSearchNet dataset provides the model with general knowledge about programming and language syntax.
However, to reason about a bug in the source code, the model should be able to differentiate between buggy and bug-free code.
To equip the model with such a reasoning capability, we use diffSBT sequences of both buggy and bug-free AST nodes (Step 4, Fig.~\ref{subfig: schema-training}).
We pre-train the Bugsplainer model to predict commit messages from the diffSBT sequences of the buggy and bug-free code.
We refer to this pre-training step as \emph{discriminatory pre-training} since Bugsplainer learns to discriminate between buggy and bug-free code.
The diffSBT sequences for the buggy and bug-free code are separated by a special token (\texttt{</s>}).
We hypothesize that the model can differentiate and attend to (i.e., selectively focus on) the changes in both sides of the separator token and generate the commit message (a.k.a., bug explanation) accordingly.
Our experimental result reports the effectiveness of discriminatory pre-training in explaining software bugs (see RQ\textsubscript{2} in Section~\ref{subsec: evaluation}).

\subsubsection{Fine-tuning}
\label{sssec: fine-tuning}

Once the discriminatory pre-training is complete, we also train Bugsplainer to generate explanations from only buggy code.
We take diffSBT sequences of only buggy code as the input and corresponding explanation (i.e., commit message) as the output.
We pass both input and output to the RoBERTa tokenizer.
Then, we fine-tune our pre-trained model from the previous phase to generate explanations from the diffSBT sequence of buggy code (Step 5, Fig.~\ref{subfig: schema-training}).
The output of the fine-tuning step is the Bugsplainer model for bug explanation generation.

\subsection{Generate Explanation}
\label{subsec: explanation-generation}
Once the training phase is complete, we test our model using the testing instances.
Fig.~\ref{subfig: schema-testing} shows how Bugsplainer generates an explanation from buggy code.
During the generation phase, Bugsplainer takes two inputs -- the buggy code and the line numbers within the code that need an explanation.
From the buggy code, Bugsplainer constructs the AST (Step 1, Fig.~\ref{subfig: schema-testing}) and extracts the AST nodes that intersect with the given line numbers (Step 2, Fig.~\ref{subfig: schema-testing}).
Subsequently, Bugsplainer converts the intersecting nodes into a diffSBT sequence (Step 3, Fig.~\ref{subfig: schema-testing}).
Then, it tokenizes the diffSBT sequence using the same RoBERTa tokenizer and passes the tokens to the fine-tuned model (Step 4, Fig.~\ref{subfig: schema-testing}).
Finally, the fine-tuned model generates an explanation for the buggy code.

\section{Experiment}
\label{sec: experiment}

We curate a large dataset of $\approx$~150K bug-fix commits and evaluate Bugsplainer using three appropriate metrics from the relevant literature -- BLEU score~\parencite{papineni2002bleu}, Semantic Similarity~\parencite{haque2022semantic}, and Exact Match.
To place our work in the literature, we compare our solution -- Bugsplainer -- with four relevant baselines.
We also conduct a developer study to assess the quality of our automatically generated explanations (e.g., accuracy, usefulness) for software bugs.
In our experiments, we thus answer four research questions as follows.

\begin{itemize}
    \item \textbf{RQ\textsubscript{1}}: How does Bugsplainer perform in explaining software bugs in terms of automatic evaluation metrics?

    \item \textbf{RQ\textsubscript{2}}: How do (a) structural information and (b) discriminatory pre-training influence the performance of Bugsplainer in generating explanations for software bugs?
    
    \item \textbf{RQ\textsubscript{3}}: Can Bugsplainer outperform the existing baseline techniques in terms of automatic evaluation metrics?
    
    \item \textbf{RQ\textsubscript{4}}: How accurate, precise, concise, and useful are the explanations of Bugsplainer compared to baselines?
\end{itemize}

\subsection{Dataset Construction}
\label{subsec: dataset-construction}

To conduct our experiments, we curate a dataset of $\approx$~150K bug-fix commits from GitHub\footnote{Accessed: April 18, 2022} using its REST API\footnote{https://docs.github.com/en/rest}. 
We discuss different steps of dataset construction as follows.

\subsubsection{Repository Selection}
First, to ensure high-quality commits, we aim to find $\approx 10K$ Python repositories with high star counts.
We choose Python since it is the second most popular programming language according to StackOverflow survey 2021\footnote{https://bit.ly/3cmooLv}. 
As GitHub's search API does not return more than $1K$ results from a single query, we use small buckets of star counts to renew our query contents.
We found the 10,000$^{th}$ repository falls in the bucket of $300\mhyphen 399$ stars.
Thus, we collect all the repositories that have a star count of $\ge 300$, which led us to a total of $10{,}154$ repositories.

\subsubsection{Collection of Bug-fix Commits}

We collect all the commits from the above repositories, which led to a total of $\approx$~11.8M commits.
Then, we attempt to find the bug-fix commits from them.
Similar to previous studies~\parencite{fischer2003populating, tufano2019empirical}, we consider a commit as a \emph{bug-fix} commit if it contains either `fix' or `solve' in its commit message.
This filtration step led us to a total of $\approx$~1.4M bug-fix commits.

\subsubsection{Filtration of Noisy Commits}
\label{subsec: dataset-construction-noiseFilteration}
To ensure commit quality, we perform a manual analysis of 500 commits that were randomly sampled from the above commit collection.
We found seven machine-generated templates in the commit messages (e.g., ``Merge branch X to master'') that can be easily detected using appropriate regular expressions (see replication package for details).
We remove these machine-generated templates from commit messages.
If a commit message contains only machine-generated texts, then the whole commit is discarded from the dataset. 
We also note that Python repositories contain 
non-Python files (e.g., configuration files) and test scripts in their commits, which are out of the scope of this work.
We thus keep the commits that have at least one modified Python file (excluding test scripts) in them.

\textcite{vaswani2017attention} report that the complexity of transformer models increases quadratically with the length of input and output sequences.
Therefore, we set a limit to the maximum length of both commit diff and commit message.
In particular, we retain such commits that have $\le$30 tokens in their commit message and $\le$170 tokens in their commit diff.  
These limits cover $>$85\% of both commit messages and commit diffs from the $\approx$~1.4M bug-fix commits.
Then, we remove commits with less than five tokens in their messages to discard trivial commits.
We also keep only the commits that have one diff hunk (i.e. change location) to avoid tangled commits (i.e. commits doing more than one task).
After performing all these noise filtration steps, we end up with $\approx$~180K bug-fix commits.

To determine the reliability of our constructed dataset, we perform a manual analysis using 385 commits.
We randomly sample these commits from $\approx$~180K commits above with a 95\% confidence level and 5\% error margin.
We find that 92.1\% of these commits are bug-fix and 5.2\% are style-fix, which indicates a negligible amount of noise in our constructed dataset. 
Previous studies~\parencite{fischer2003populating, tufano2019empirical} also use datasets with similar amount of noise.
Furthermore, manually filtering $\approx$~180K commits was prohibitively costly or impractical, which possibly justifies our choice of using the current version of the dataset.

\subsubsection{Embedding Structural Information}
\label{sssec: embed-struct}

We generate diffSBT sequence for each commit as described in Section~\ref{subsec: gen-diffsbt}.
We first generate AST for both the buggy and bug-free code from the commit using the \texttt{ast} parser of Python 3\footnote{https://docs.python.org/3/library/ast.html}. 
After discarding syntactically incorrect programs, we find a total of $\approx$~150K diffSBT sequences.

\subsubsection{Construction of Training and Testing Data}
\label{sssec: const-splits}

First, we randomly select 110K entries as the training dataset for both pre-training and fine-tuning steps.
Second, we randomly split the remaining 40K entries into four sets that are allocated for validation and testing in both pre-training and fine-tuning steps.
Thus, the training data are shared by both pre-training and fine-tuning steps whereas the validation and testing data are not shared.
Finally, we remove the part of the diffSBT sequence that corresponds to the fixed version of the source code from the three fine-tuning splits (training, validation and testing) since Bugsplainer aims to generate an explanation from the buggy code only.

\subsection{Evaluation Metrics}
\label{subsec: eval-metric}

To evaluate the explanations generated by Bugsplainer, we use three different metrics -- BLEU score~\parencite{papineni2002bleu}, Semantic Similarity~\parencite{haque2022semantic}, and Exact Match.
Relevant studies~\parencite{vaswani2017attention, raffel2020exploring, wu2016google, wang2021codet5} frequently used these metrics, which justifies our choice. 
They are defined as follows.

\subsubsection{BLEU: Bi-Lingual Evaluation of Understanding}
BLEU score~\parencite{papineni2002bleu} is a widely used performance measure for NMT.
It has been used in software engineering context as well~\parencite{iyer2016summarizing, hu2018deep, jiang2017automatically, liu2018neural, tao2021evaluation, liu2020atom}.
It calculates the similarity between auto-generated and reference sequences in terms of their n-grams precision as follows.
\begin{equation}
    BLEU = BP \cdot exp\left(\sum_{n=1}^N{w_n log (p_n)}\right)
\end{equation}
Here, $p_n$ is the ratio between overlapping n-grams (from both generated and reference sequences) and the total number of n-grams in the generated sequence, and $w_n$ is the weight of the n-gram length.
Following the existing studies~\parencite{hu2018deep, tao2021evaluation}, we use $N = 4$ and $w_n = 0.25$ for all $n \in [1, N]$.
That is, we compute the mean BLEU score for all n-gram lengths.
The brevity penalty, $BP$, lowers the BLEU score if the generated sequence is too small.

There exist several variations of the BLEU score.
In our study, we use case-insensitive BLEU score with \emph{add one smoothing}~\parencite{lin2004automatic} which aligns the most with human judgement~\parencite{tao2021evaluation}.

\subsubsection{Semantic Similarity}
Although the BLEU score is widely adopted for evaluating machine translation, it does not take the meaning of the text into account.
\textcite{haque2022semantic} conduct a human study to determine which metric better represents the perception of human evaluators.
They find that Sentence-BERT encoder~\parencite{reimers2019sentence} with cosine similarity has the highest correlation with the human evaluated similarity.
Sentence-BERT provides a fixed-length numeric representation for any given text.
As suggested by \textcite{haque2022semantic}, we use \texttt{stsb-roberta-large}\footnote{https://bit.ly/3dR9mxD} pre-trained model to generate the embedding for the input text.
We compute the Semantic Similarity as follows. 
\begin{equation}
    SemSim(ref, gen) = {cos(sbert(ref), sber t(gen))}
\end{equation}
where $sbert(x)$ is the numerical representation from Sentence-BERT for any input text $x$, $ref$ is the reference explanation, and $gen$ is the generated explanation.

\subsubsection{Exact Match}
We also use the Exact Match metric to evaluate our explanation.
As the name suggests, Exact Match checks whether a generated explanation exactly matches the corresponding reference explanation.
It is analogous to string equality check in many programming languages, which is case-sensitive and space-sensitive.

\subsection{Evaluating Bugsplainer}
\label{subsec: evaluation}

\textbf{Answering RQ\textsubscript{1} -- Performance of Bugsplainer}:
Table~\ref{tab: rq1} shows the performance of Bugsplainer in terms of BLEU score, Semantic Similarity, and Exact Match.

When the dataset is split randomly into training, validation and testing sets, Bugsplainer achieves a BLEU score of 33.15, which is considered as \emph{understandable} to \emph{good} translation according to Google's AutoML Translation documentation\footnote{https://bit.ly/3wGpCIx}.
Explanations from Bugsplainer also have an average of 55.76\% Semantic Similarity, which indicates a major semantic overlap with the explanations from developers.
Finally, 22.37\% of the explanations exactly match the reference explanations.
To achieve an Exact Match with the reference, an NMT model warrants a substantial knowledge of the domain.
All these statistics are highly promising and demonstrate the high potential of our technique in explaining software bugs.

\textcite{allamanis2019adverse} report that an overlap between training and testing datasets might lead to an overestimation in performance measurement.
In our experiment design, we ensure that there is no overlap between our training and testing datasets (see Section~\ref{sssec: const-splits}).
However, we also use a pre-trained CodeT5~\parencite{wang2021codet5} model which is pre-trained on millions of code snippets from thousands of repositories in CodeSearchNet dataset~\parencite{husain2019codesearchnet}.
As a result, there might be an unavoidable overlap between pre-training and testing datasets.
To ensure a fair evaluation, we thus discard the testing instances from CodeSearchNet repositories ($\approx$~14\% instances) to avoid any possibility of overlap and re-evaluate Bugsplainer.
Table~\ref{tab: rq1} shows that after discarding the overlapping repositories, Bugsplainer demonstrates a marginal improvement both in BLEU score and Semantic Similarity.

\begin{table}
\centering
\caption{Performance of Bugsplainer}
\label{tab: rq1}
\begin{tabular}{|l|p{22mm}|l|p{12.25mm}|p{8mm}|}
\hline
\textbf{Model}    & \textbf{Dataset} & \textbf{BLEU} & \textbf{Semantic Similarity} & \textbf{Exact Match} \\ \hline \hline
\multirow{3}{14mm}{
Bugsplainer}      & Random split     & 33.15         & 55.76                        & 22.37                \\ \cline{2-5}
                  & No CodeSearchNet Repository    & 34.53         & 56.67                        & 19.55                 \\ \cline{2-5}
                  & Cross-project    & 17.16         & 44.98                        & 7.15                 \\ \hline \hline
\multirow{3}{14mm}{
Bugsplainer 220M} & Random split     & 33.87         & 56.35                        & 23.50                \\ \cline{2-5}
        & No CodeSearchNet Repository    & 35.59         & 57.29                        & 20.74                 \\ \cline{2-5}
                  & Cross-project    & 23.83         & 49.00                        & 15.47                \\ \hline
\end{tabular}
\end{table}

In the real world, when adopting Bugsplainer for a new project, data from the new project might not always be available to re-train Bugsplainer.
Therefore, we investigate how the performance of Bugsplainer varies in a cross-project setting.
That is, each of the training, validation, and testing datasets contain commits from mutually exclusive projects.
From Table~\ref{tab: rq1}, we see that even though the performance of Bugsplainer decreases in a cross-project setting, it is still promising, especially in terms of the Semantic Similarity metric.
We see that the BLEU score decreases by 48\% whereas the Semantic Similarity decreases by only 19\%.
That is, in the cross-project setting, the generated explanations might express similar information but with different words.
To verify the case, we manually compare a sample of 385 explanations from Bugsplainer (95\% confidence level and 5\% error margin) with the reference explanations.
We find that a substantial amount of generated explanations express information either more precisely or with different phrases, which might cause the BLEU score to be low.
For instance, for a particular bug\footnote{https://bit.ly/3RoSpsT}, Bugsplainer generates \textit{``Improve the message in IncompleteRead.\_\_init\_\_"}, whereas the reference is \textit{``fixing incorrect message for IncompleteRead."}
Even though the generated explanation is accurate and more precise, it returns a BLEU score of only 11.
Such a phenomenon also explains the low BLEU score and comparatively high Semantic Similarity score for the cross-project setting of Bugsplainer.

Recent studies suggest that increasing the model size can significantly improve the performance of deep learning models~\parencite{raffel2020exploring, liu2019roberta, wang2021codet5}.
We thus were interested to see how the performance of Bugsplainer changes with an increased number of parameters.
For this experiment, we train a 220M parameter variant of Bugsplainer and call it \emph{Bugsplainer 220M}.
Both variants share the same architecture (i.e., T5) but they have different hyperparameters.
The detailed hyperparameter values can be found in the replication package.
Table~\ref{tab: rq1} also shows the performance of Bugsplainer 220M in the random split and cross-project settings.
We see improved performance in both cases, which aligns with the existing findings~\parencite{raffel2020exploring, wang2021codet5}.
Interestingly, in cross-project settings, Bugsplainer 220M achieves a big bump of $\approx$~39\% in BLEU score and $\approx$~117\% in Exact Match.
Such a finding suggests that Bugsplainer 220M can generalize the acquired knowledge better across multiple projects than Bugsplainer.

\begin{table}
  \centering
  \caption{Performance of Bugsplainer by Input Length}
  \label{tab: score-by-input-len}
  \begin{tabular}{|c|c|c|c|}
    \hline
    \textbf{\#Words} & \textbf{BLEU} & \textbf{Semantic Similarity} & \textbf{Exact Match} \\ \hline
    \# $< 50$ & 32.05 & 54.90 & 17.84 \\ \hline
    $50 \leq$ \# $< 100$ & 34.22 & 56.25 & 18.65 \\ \hline
    $100 \leq$ \# $< 150$ & 34.72 & 56.99 & 21.10 \\ \hline
    $150 \leq$ \# $< 200$ & 32.92 & 56.50 & 23.91 \\ \hline
  \end{tabular}
\end{table}

\begin{table}
  \centering
  \caption{Performance of Bugsplainer by The Length of Ground Truth}
  \label{tab: score-by-output-len}
  \begin{tabular}{|c|c|c|c|}
    \hline
    \textbf{\#Words} & \textbf{BLEU} & \textbf{Semantic Similarity} & \textbf{Exact Match} \\ \hline
    \# $< 10$ & 35.75 & 56.32 & 22.72 \\ \hline
    $10 \leq$ \# $< 20$ & 27.70 & 53.16 & 8.93 \\ \hline
    $20 \leq$ \# $< 30$ & 21.62 & 52.03 & 1.26 \\ \hline
  \end{tabular}
\end{table}

Finally, we investigate how the performance of Bugsplainer is affected by the input and output length.
Table~\ref{tab: score-by-input-len} shows the metric scores categorized by the number of words in the input buggy code segments.
The table shows no clear correlation between the input length and the performance.
With increasing input length, the performance both increases and decreases.
On the contrary, Table~\ref{tab: score-by-output-len} shows that the performance of Bugsplainer tends to decrease with increasing ground truth lengths.
Interestingly, the drop in Semantic Similarity is not as strong as the BLEU score or Exact Match score.
This suggests that even with increasing output length, Bugsplainer can provide explanations that are semantically coherent with the ground truth.

\vspace{2mm}
\noindent
\fcolorbox{black}{gray!20}{
\parbox{.95\linewidth}{
\textbf{Summary of RQ\textsubscript{1}:} Bugsplainer can generate bug explanations that are \emph{understandable} and \emph{good} according to Google's standard. It shows promising results not only in random split settings but also in cross-project settings. With a higher number of parameters, Bugsplainer can better generalize the acquired knowledge across multiple projects. }}
\indent

\textbf{Answering RQ\textsubscript{2} -- Role of structural information and discriminatory pre-training in Bugsplainer}:
In this experiment, we analyze the impact of structural information and discriminatory pre-training on bug explanation generation.
We remove one of these two components from Bugsplainer and keep the rest as is.
Such an experiment helps us understand the contribution of individual components toward Bugsplainer.

To analyze the impact of structural information, we use raw commit diff as input rather than diffSBT sequences.
In the pre-train dataset, we keep the commit diff as is, while in the fine-tuning dataset, we remove the added lines (i.e., bug-free lines) from the commit diff.
From Table~\ref{tab: rq2}, we see that the BLEU score of Bugsplainer reduces by 7.15\% due to the absence of structural information. Interestingly, the Exact Match score also drops by 31.07\%, which is significant.

To analyze the impact of discriminatory pre-training, we use only fine-tuning dataset and avoid the pre-training step.
In this experiment, we use the diffSBT sequences as input during the fine-tuning step.
From Table~\ref{tab: rq2}, we see that the BLEU score of Bugsplainer reduces by 8.54\% due to the absence of discriminatory pre-training. Interestingly, the Exact Match score also drops by 25.70\%, which is significant. 

The significant performance drops due to the absence of structural information and discriminatory pre-training indicate their important roles in Bugsplainer. Our technique also performs the best when both items are incorporated.

\begin{table}
\centering
\caption{Role of structural information and discriminatory pre-training}
\label{tab: rq2}
\begin{tabular}{|p{32mm}|l|p{13mm}|l|}
\hline
\textbf{Model}                                & \textbf{BLEU}   & \textbf{Semantic Similarity} & \textbf{Exact Match} \\ \hline \hline
\textbf{Bugsplainer}                          & \textbf{33.15}  & \textbf{55.76}               & \textbf{22.37}       \\ \hline
Bugsplainer without \newline structural information       & 30.78           & 53.74                        & 15.42                \\ \hline
Bugsplainer without \newline discriminatory pre-training & 30.32           & 53.51                        & 16.62                \\ \hline
\end{tabular}
\end{table}

\vspace{2mm}
\noindent
\fcolorbox{black}{gray!20}{
\parbox{.95\linewidth}{
\textbf{Summary of RQ\textsubscript{2}:} Both structural information and discriminatory pre-training have a major contribution to the performance of Bugsplainer. Furthermore, they are the most effective when they are used together.}
}
\indent

\textbf{Answering RQ\textsubscript{3} -- Comparison with existing baseline techniques}:
In this research question, we compare Bugsplainer with existing techniques from the literature and investigate whether Bugsplainer can outperform them in terms of various evaluation metrics.
To the best of our knowledge, there exists no work that explains software bugs in natural language texts. 
Since commit message generation is quite similar to explanation generation, we use state-of-the-art commit message generation techniques as our baseline.
The main difference between commit message generation and explanation generation is the former takes both buggy and bug-free lines as input whereas the latter takes only buggy lines as input.
In particular, we compare Bugsplainer with three commit message generation techniques namely -- \emph{CommitGen}~\parencite{jiang2017automatically}, \emph{NNGen}~\parencite{liu2018neural}, and \emph{Fine-tuned CodeT5}~\parencite{wang2021codet5} and a static analysis tool \emph{pyflakes}~\parencite{pyflakes}.
None of these existing approaches for commit message generation learns to differentiate between buggy and bug-free code. Thus, our approach has a better chance of generating meaningful explanations for the buggy code.


To generate error messages from \emph{pyflakes}, a static analysis tool, we run pyflakes on the whole buggy source code. 
Once we have the error messages, we keep only the messages generated for the buggy lines. If we get multiple errors for the same data point, we keep them all and report the one with the highest automatic metric score (e.g., BLEU score).

CommitGen uses an NMT framework namely nematus~\parencite{sennrich2017nematus}, which we use for our replication of the technique.
The authors also provide the values of all the important parameters in their paper, which were carefully adopted in our replication. 

According to a recent study~\parencite{tao2021evaluation}, NNGen~\parencite{liu2018neural} is the state-of-the-art tool for generating commit messages.
Being an Information Retrieval-based technique, NNGen does not require any training phase.
It solely depends on the K-Nearest Neighbours algorithm.
NNGen first finds k most similar commit diffs from the training set using \emph{bag-of-words} model (i.e., term frequency) and cosine similarity measure.
Since the authors do not provide any details of their bag-of-words implementation, we use the \texttt{CountVectorizer} API of the scikit-learn library in our replication.
As suggested in the paper, we set the value of k to 5.
From the top-k commits, NNGen selects the message from the most similar commit (using the BLEU score) as the final translation.

The fine-tuned CodeT5 model has the same architecture and the same hyperparameters as those of Bugsplainer.
However, unlike Bugsplainer, it neither uses the structural information from the source code nor performs any discriminatory pre-training.

\begin{table}
\centering
\caption{Comparison of Bugsplainer with existing baseline techniques (Using five random runs)}
\label{tab: rq4}

\begin{tabular}{|l|l|l|l|}
\hline
\textbf{Technique}       & \textbf{BLEU}  & \textbf{Semantic Similarity} & \textbf{Exact Match} \\ \hline \hline
pyflakes             & 0.49           & 5.68                         & 0.00                 \\ \hline
CommitGen            & 9.94           & 35.39                        & 1.04                 \\ \hline
NNGen                & 24.16          & 47.33                        & 14.17                \\ \hline
Fine-tuned CodeT5    & 26.19          & 54.52                        & 8.85                 \\ \hline
\textbf{Bugsplainer}$^*$
& \textbf{32.90} & \textbf{55.22}               & \textbf{18.14}       \\ \hline
\end{tabular}
\footnotesize{
\newline \newline
$^*$The scores differ from the earlier tables due to five random runs}
\vspace{1mm}
\end{table}

Table~\ref{tab: rq4} shows the comparison between Bugsplainer and four baselines in terms of BLEU score, Semantic Similarity, and Exact Match.
The results shown in the table are the mean of five runs with different random initialization of the parameters.
We see that Bugsplainer outperforms all the baselines in terms of all three metrics.
Only Fine-tuned CodeT5 is comparable with Bugsplainer.
Therefore, we perform the Mann-Whitney U rank test~\parencite{mann1947test} to see whether their performances over the five runs are significantly different.
We found that Bugsplainer performs significantly higher than Fine-tuned CodeT5 i.e., \emph{p-value} = 0.008 $<$ 0.05, Cliff's \emph{d} = 1.0 (\emph{large}) for all three metrics.

CommitGen relies on certain patterns in commit messages that might be generated by machines~\parencite{liu2018neural}. 
However, we removed auto-generated messages to ensure high-quality dataset (Section~\ref{subsec: dataset-construction-noiseFilteration}).
CommitGen is also based on the LSTM architecture that does not perform well with long inputs~\parencite{vaswani2017attention}.
Thus, the low scores of CommitGen are explainable.

According to Google’s AutoML Translation documentation, a BLEU score between 20 and 29 indicates that the gist of a generated message is clear, but has significant errors.
NNGen reuses existing commit messages from the training set and thus cannot analyze the dynamic behaviour of software programs.
Thus, such errors in the translation are also explainable.

\textcite{thung2015extent} report that static analysis tools produce a lot of false negatives.
This means they do not produce any output for potentially buggy code in many cases.
Our experiment with \emph{pyflakes}, shows a similar result, generating error messages for only 7.70\% cases.
Therefore, its poor metric scores are also understandable.

\vspace{2mm}
\noindent
\fcolorbox{black}{gray!20}{
\parbox{.95\linewidth}{
\textbf{Summary of RQ\textsubscript{3}:} Bugsplainer outperforms all four baselines in terms of three performance metrics. According to our statistical tests, our technique outperforms the closest competitor -- Fine-tuned CodeT5 -- by a statistically significant margin. }
}
\indent

\textbf{Answering RQ\textsubscript{4} -- Evaluation of Bugsplainer using a developer study}:
The metric-based evaluation demonstrates the benefit of our technique in bug explanation generation.
We also conduct a developer study to further demonstrate the benefit of Bugsplainer in a practical setting.
Given the reference explanations of a software bug (e.g., the message of a bug-fix commit), we ask the developers to assess how accurate, precise, concise, and useful the explanations are.
During the study, \emph{we anonymize the model names} to avoid any bias.

\begin{table}
    \caption{Quality aspects of generated explanations}
    \label{tab: defintions}
    \centering
    \begin{tabular}{|l|p{65mm}|}
        \hline
        \textbf{Quality} & \textbf{Overview} \\ \hline
        \hline
        \emph{Accurate} & It provides the same factual information as the reference. \\ \hline
        \emph{Precise} & It can pinpoint the issue in the code. \\ \hline
        \emph{Concise} & It is short and still conveys the whole message. \\ \hline
        \emph{Useful} & The provided information has the potential to fix the bug. \\ \hline
    \end{tabular}
\end{table}

\begin{figure*}
    \centering
    \includegraphics[width=\linewidth]{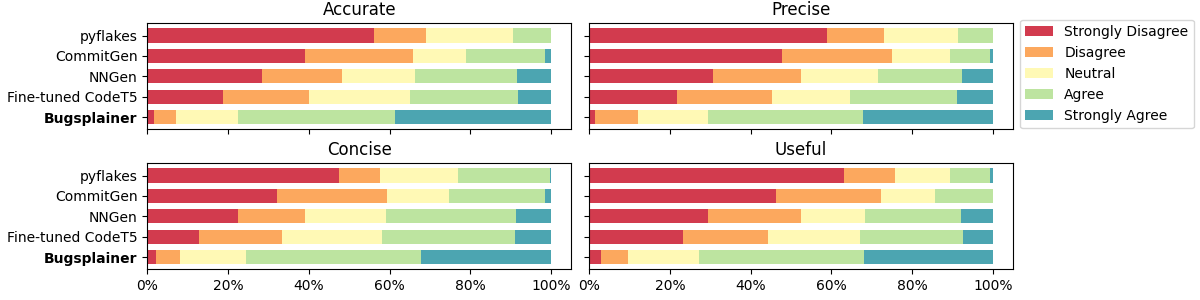}
    \caption{Comparison of Bugsplainer with the baselines using Likert scores}
    \label{fig: dev-study}
\end{figure*}

\textbf{Study participants}:
The target population of our study is English-speaking software engineers with experience in Python programming language.
We invite our participants in two ways.
First, we contact software companies with a history of participation in academic studies to contribute to this research.
Second, we advertise the study on the authors' social networks to reach potential participants and increase the diversity of samples.
As of August 31, 2022, we receive a total of 20 responses to our developer study.
The participants have professional software development experience of 1 to 10 years and bug-fixing experience of 1 to 7 years.
All of them are familiar with Python programming language as well.
Such experience makes them suitable candidates for our study.

\textbf{Study setup}:
In the developer study, each participant worked with 15 bug-fix commits and spent 30 minutes on average.
To select these examples for our developer study, we apply random sampling without replacement to the testing set.
To avoid information overload, we take the examples that (1) do not have more than five changed lines or more than 15 word-tokens in a single line within the commit diff, and (2) do not require any project-specific knowledge to understand the bug.
We take the first 15 randomly sampled examples matching these two criteria.

We ask the participants to assess the accuracy, precision, conciseness, and usefulness of the explanations from Bugsplainer and baselines with respect to the reference explanations.
Table~\ref{tab: defintions} provides our definitions for these aspects.
The participants assess these four aspects using a five-point Likert scale, where 1 indicates strongly disagree and 5 indicates strongly agree.
Please note that \emph{we anonymize the model names and do not show the participants which explanation comes from which model to avoid any potential bias}.
We collect a total of 300 data points (15 questions $\times$ 5 explanations $\times$ 4 aspects) from each participant.

\textbf{Study result and discussion}:
Table~\ref{tab: dev-study} summarizes our findings from the developer study.
We note that the participants find the explanations from Bugsplainer to be the most \emph{accurate}, most \emph{precise}, most \emph{concise} and most \emph{useful}.
Based on the median and mode values, we see that the participants agree the most with explanations from Bugsplainer.
Similar to our findings in RQ\textsubscript{3}, according to the developers, the closest competitor of Bugspaliner is Fine-tuned CodeT5.
According to the mode values, the developers agree with Fine-tuned CodeT5 in many cases.
However, looking at the 2$^{nd}$ mode values, we see that the developers strongly disagree with Fine-tuned CodeT5 in a noticeable manner, making the mean agreement poor.
We also perform the Wilcoxon Signed Rank test to check whether the developers' agreement with Bugsplainer is significantly higher than that of Fine-tuned CodeT5.
For accuracy, conciseness, precision and usefulness, the p-values are $5.16e{-23}$, $2.74e{-17}$, $7.45e{-18}$ and $2.63e{-23}$ respectively, all are below the threshold of $0.05$, which makes the difference significant.

Fig.~\ref{fig: dev-study} shows the distribution of participants' agreement levels in different aspects.
We see that the participants disagree with Bugsplainer very few times (highest 14\% times in precision) with a substantial amount of agreement (highest 76\% in accuracy).
Nearly half of the time the developers strongly disagree with pyflakes and CommitGen.
Such disagreement with pyflakes is explicable since it does not generate any error message for 13 out of 15 cases.
However, CommitGen, even after explaining all cases, receives a high disagreement due to its generic and less informative explanations.


\vspace{2mm}
\noindent
\fcolorbox{black}{gray!20}{
\parbox{.95\linewidth}{
\textbf{Summary of RQ\textsubscript{4}:} Professional developers with bug-fixing experience find the bug explanations from Bugsplainer to be accurate, precise, concise, and useful. Their preference levels for Bugsplainer over other baseline techniques are also significantly higher.}
}
\indent

\section{Related Work}
\label{sec: related-works}

\subsection{Explanation of Software Bugs}
Software bugs are errors, flaws, or defects in a program 
that prevent the program from working correctly \parencite{ieee-standard}.
They claim 50\% of development time and cost the global economy billions of dollars every year
\cite{rahman2020some}.
While there have been numerous approaches to find or repair software bugs, neither many approaches attempt to explain the bugs in the source code to the developers, nor are they practical and scalable enough for industry-wide use~\parencite{kochhar2016practitioners, zou2018practitioners}.
Several tools (e.g. FindBugs~\parencite{FindBugs}, PMD, SonarLint, PyLint, and pyflakes~\parencite{pyflakes}) attempt to explain bugs using static analysis.
Unfortunately, their utility could be limited due to their high false-positive results and lack of meaningful, actionable explanations~\parencite{barik2016should, johnson2013don, ayewah2007evaluating}.
Furthermore, their explanations can be too generic and limited by their templated natures~\parencite{nachtigall2022large}.
Recent studies suggest complementing these messages with rule graphs~\parencite{do2020explaining}, assertive error explanation~\parencite{barik2016should}, and interactive feedback from developers~\parencite{ko2004designing}.

\begin{table}
\centering
\caption{Comparison of Bugsplainer with baselines using developer study}
\label{tab: dev-study}
\begin{tabular}{|l|l|l|l|l|p{6.5mm}|} \hline
\textbf{Quality}    & \textbf{Model} & \textbf{Mean} & \textbf{Median} & \textbf{Mode} & \textbf{2$^{nd}$ Mode} \\ \hline \hline
\multirow{5}{*}{Accurate}                                    & pyflakes     & 1.841 & 1 & 1 & 3 \\\cline{2-6} 
                               & CommitGen    & 2.176 & 2 & 1 & 2 \\\cline{2-6} 
                               & NNGen        & 2.653 & 3 & 1 & 4 \\\cline{2-6} 
                               & Fine-tuned CodeT5 & 2.842 & 3 & 4 & 3 \\\cline{2-6} 
 & \textbf{Bugsplainer} & \textbf{4.074} & \textbf{4} & \textbf{5} & \textbf4 \\\hline 
\multirow{5}{*}{Precise}                                     & pyflakes     & 1.768 & 1 & 1 & 3 \\\cline{2-6} 
                               & CommitGen    & 1.884 & 2 & 1 & 2 \\\cline{2-6} 
                               & NNGen        & 2.529 & 2 & 1 & 2 \\\cline{2-6} 
                               & Fine-tuned CodeT5 & 2.772 & 3 & 4 & 2 \\\cline{2-6} 
 & \textbf{Bugsplainer} & \textbf{3.891} & \textbf{4} & \textbf{4} & \textbf5 \\\hline 
\multirow{5}{*}{Concise}                                     & pyflakes     & 2.182 & 2 & 1 & 4 \\\cline{2-6} 
                               & CommitGen    & 2.350 & 2 & 1 & 2 \\\cline{2-6} 
                               & NNGen        & 2.881 & 3 & 4 & 1 \\\cline{2-6} 
                               & Fine-tuned CodeT5 & 3.044 & 3 & 4 & 3 \\\cline{2-6} 
 & \textbf{Bugsplainer} & \textbf{3.974} & \textbf{4} & \textbf{4} & \textbf5 \\\hline 
\multirow{5}{*}{Useful}                                      & pyflakes     & 1.724 & 1 & 1 & 3 \\\cline{2-6} 
                               & CommitGen    & 1.960 & 2 & 1 & 2 \\\cline{2-6} 
                               & NNGen        & 2.576 & 2 & 1 & 4 \\\cline{2-6} 
                               & Fine-tuned CodeT5 & 2.728 & 3 & 4 & 1 \\\cline{2-6} 
 & \textbf{Bugsplainer} & \textbf{3.923} & \textbf{4} & \textbf{4} & \textbf5 \\\hline 
\end{tabular}
\end{table}

Besides the static analysis, there have been several attempts to explain a program's behaviours, failed tests, bug-fixing patches, undocumented code, and intelligent behaviours.
\textcite{ko2004designing} design an interrogative debugging system for the Alice programming environment~\parencite{Alice} where a novice learner can inquire why a program behaves unexpectedly or why it does not show an expected behaviour.
\textcite{lim2009and} later suggest that these why and why not questions are essential to improve a user's understanding or perception of an intelligent system.
\textcite{zhang2011automated} explain a failed test case by automatically performing failure-correcting edits (e.g., replacement of identifiers' values) and synthesizing a code comment from them.
\textcite{befrouei2016abstraction} use program execution traces to explain concurrency bugs. 
Later \textcite{bragaglio2021system} remove irrelevant information from the execution traces to understand the cause of the unexpected behaviour.
\textcite{liang2019explain} investigate what should be included in a patch explanation, such as expected program behaviours or a high-level summary of code changes.
Recently, \textcite{pornprasitpyexplainer} also explain why the changed code can be defect-prone by visualizing how specific local rules are satisfied by a change.
Although all these studies and approaches are relevant and are a source of our inspiration, they might be restricted to only specific problem contexts (e.g., Alice~\parencite{Alice}, failed tests~\parencite{zhang2011automated}) or certain types of bugs (e.g., concurrency bugs~\parencite{befrouei2016abstraction}).

Unlike these traditional approaches, Bugsplainer is not restricted to any specific context or bugs. Besides, it 
can generate explanations that resemble that of humans, and are accurate, precise, concise, and useful (check RQ$_4$ for details). 

\subsection{Translation of Source Code into Texts}
Our work is also related to code translation into natural language texts. Existing approaches translate code into the natural language to generate code comments, review comments, and commit messages.

Earlier works on code comment generation utilize hand-crafted templates~\parencite{mcburney2014automatic, sridhara2011automatically} and information retrieval~\parencite{haiduc2010use, wong2013autocomment}, while recent works more depend on learning-based approaches~\parencite{allamanis2016convolutional, iyer2016summarizing, hu2018deep}.
\textcite{wei2020retrieve} combine both IR and NMT in comment generation.
Recently, \textcite{mastropaolo2021studying} use the Text-To-Text Transfer Transformer (T5) to perform several tasks including code comment generation.
Their comments explain what is happening in the code rather than what makes the code buggy. 
On the contrary, Bugsplainer not only explains the buggy code but also suggests useful information to correct the bug in the code (e.g., Table \ref{tab: generated-explnations}).

To generate code review comments, \textcite{tufano2021towards} make use of the CodeT5~\parencite{wang2021codet5} model with Stack Overflow data and fine-tune their model with pairs of function and review comment pairs from GitHub.
Later, \textcite{hong2022commentfinder} use Gestalt Pattern Matching (GPM) to mine candidate review comments of similar methods from a large corpus of source code repositories. 
However, both approaches treat source code as regular texts (e.g., a sequence of tokens) overlooking the structures. 
On the other hand, Bugsplainer leverages the structures of code through diffSBT sequences and learns to differentiate between buggy and bug-free code as a part of explanation generation.

To generate commit messages, several studies adopt an attention mechanism with RNN~\parencite{jiang2017automatically, loyola2017neural, xu2019commit, liu2020atom} and leverage structural information~\parencite{xu2019commit, liu2020atom}.
\textcite{xu2019commit} jointly model the semantic representation and structural representation of code changes where they substitute identifier names with placeholders in the code.
\textcite{liu2020atom} capture both ASTs of code changes where they convert each AST into path sequence. 
However, the conversion of each change from the AST into its own path might lead to long, redundant sequences, which could hurt the translation performance.
\textcite{liu2018neural} show that a simple information retrieval-based approach, NNGen, has promising capability in commit message generation due to its repetitive nature.
Both techniques above are closely related to ours due to their nature of translation. We thus compare Bugsplainer with them using experiments and the details can be found in Section~\ref{subsec: evaluation}.

\section{Threats To Validity}
\label{sec: threats}

Threats of \emph{internal validity} relate to experimental errors and biases.
Re-implementation of the existing baseline techniques could pose a threat.
However, our implementation of NNGen~\parencite{liu2018neural} is based on the well-known k-nearest neighbour algorithm.
CommitGen~\parencite{jiang2017automatically} uses a framework and reports all important hyperparameters.
We use the same framework and the reported parameters for our implementation.
For pyflakes, we use the officially provided package and follow the official documentation.
Fine-tuned CodeT5 adopts the same model architecture as that of Bugsplainer. 
Thus, threats related to replication might be mitigated.
We also repeat our experiments five times and compare the performance with that of baselines to mitigate any bias due to random trials.

Threats to \emph{external validity} relate to the generalizability of our work.
Even though Bugsplainer is evaluated using only Python code, the underlying algorithm is language-agnostic and can be easily adapted to any traditional programming language.
The use of metric-based evaluation might also pose threat to the real-world usability of our approach~\parencite{tao2021evaluation, haque2022semantic}.
To mitigate this threat, we also conduct a developer study involving 20 participants from six different countries.
As the developer study suggests, our bug explanations were also found to be accurate and useful in real-world scenarios.

Finally, the performance of Bugsplainer might depend on the precision of bug localization tools.
To minimize this dependency, Bugsplainer accepts a range of lines containing both buggy lines and their surrounding lines as input during explanation generation.
However, we do not indicate which lines among them contain a bug. Thus, precise localization of the bug either by developers or by existing tools might not be necessary to generate explanations using our tool.
 
\section{Conclusion and Future Works}
\label{sec: conclusion}
Software bugs not only claim precious development time but also cost billions every year. Although there have been many approaches for finding or repairing software bugs, there exists little research on automatically explaining the bugs.
In this paper, we propose \emph{Bugsplainer}, a novel technique that generates explanations for buggy code segments.
Our technique can leverage both structural information and buggy code patterns from source code and employs neural machine translation with an attention mechanism to generate 
bug explanations.
We evaluate Bugsplainer using three metrics (i.e., BLEU score, Semantic Similarity and Exact Match) where our technique outperforms the baselines.
We also conduct a developer study involving 20 participants and our explanations were found to be more accurate and more useful compared to the baselines.

In future, we will investigate how to encode the structural information from source code in a more compact and efficient format and how to better leverage the structural differences between buggy and bug-free code. This might help us better understand the underlying semantics of software bugs.

\section*{Acknowledgement}
This work was supported by Dalhousie University and Mitacs Accelerate International Program. We would like to thank Avinash Gopal, Ben Reaves, and Massimiliano Genta from our industry partner -- \emph{Metabob Inc}. We would also like to thank all the anonymous participants in our developer study.

\scriptsize
\printbibliography

\end{document}